\pdfoutput=1

\documentclass[twocolumn]{aastex62}
\usepackage{apjfonts}
\usepackage{rotating}
\usepackage{threeparttable}
\usepackage{booktabs}
\usepackage{supertabular}
\usepackage{footnote}

\shorttitle{Statistical Study of Filament Eruptions}
\shortauthors{Zou et al.}

\begin{document}

\title{A Statistical Study of Solar Filament Eruptions That Forms
  High-Speed Coronal Mass Ejections}

\author{Peng Zou}
\affil{Institute of Space Science and Applied Technology, Harbin Institute of Technology, Shenzhen 518055, China}

\author{Chaowei Jiang$^*$}
\affil{Institute of Space Science and Applied Technology, Harbin Institute of Technology, Shenzhen 518055, China}

\author{Fengsi Wei}
\affil{Institute of Space Science and Applied Technology, Harbin Institute of Technology, Shenzhen 518055, China}

\author{Pingbing Zuo}
\affil{Institute of Space Science and Applied Technology, Harbin Institute of Technology, Shenzhen 518055, China}

\author{Yi Wang}
\affil{Institute of Space Science and Applied Technology, Harbin Institute of Technology, Shenzhen 518055, China}

\email{*Corresponding authors: chaowei@hit.edu.cn}

\begin{abstract}
  Coronal mass ejections (CMEs) play a decisive role in driving space
  weather, especially, the fast ones (e.g., with speeds above
  $800$~km~s$^{-1}$). Understanding the trigger mechanisms of fast
  CMEs can help us gaining important information in forecasting
  them. The filament eruptions accompanied with CMEs provide a good
  tracer in studying the early evolution of CMEs. Here we surveyed 66
  filament-accompanied fast CMEs to analyse the correlation between
  the trigger mechanisms, namely either magnetic reconnection or ideal
  MHD process, associated flares, and CME speeds. Based on the data
  gathering from SDO, GONG and STEREO, we find that: (1) Active region
  (AR) filament and intermediate filaments (IFs) eruptions show a
  higher probability for producing fast CMEs than quiet Sun (QS)
  filaments, while the probability of polar crown (PC) filament
  eruptions is zero in our statistic; (2) AR filament eruptions that
  produce fast CMEs are more likely triggered by magnetic
  reconnection, while QS and IFs are more likely triggered by ideal
  MHD process; (3) For AR filaments and IFs, it seems that the
  specific trigger mechanism does not have a significant influence on
  the resulted CME speeds, while for the QS filaments, the ideal MHD
  mechanism can more likely generate a faster CME; (4) Comparing with
  previous statistic study, the onset heights of filament eruptions
  and the decay indexes of the overlying field show some differences:
  for AR filaments and IFs, the decay indexes are larger and much
  closer to the theoretical threshold, while for QS filaments, the
  onset heights are higher than those obtained in previous results.
\end{abstract}


\section{Introduction} \label{intro}

Coronal mass ejections (CMEs), which are the most violent, eruptive
phenomenon on the Sun, can throws out over $10^{12} - 10^{13}$~kg of
magnetized plasma into the
interplanetary~\citep{jackson1985,gopalswamy1992,hudson1996}. The very
high-speed CMEs, if direct to the Earth, can cause strong geomagnetic
storms and severely threaten space weather
\citep{tsurutani1988,gonzalez1999,huttunen2002,huttunen2005,liu2014}. To
avoid such disaster caused by them, prediction of CMEs, especially the
fast ones, is an important topic in space weather research
\citep{singer2001,schwenn2005}, 
Particularly, it is crucial to investigate the pre-eruptive conditions
and the trigger mechanisms of CMEs for a reliable prediction.

CMEs originates from the solar corona, and are often accompanied with
other activities, such as flares and filament eruptions
\citep{harrison1995,yashiro2005,wang2007,gopalswamy2003,jing2004}. Now
it is commonly believed that these eruptive phenomena are all driven
by the evolution of solar coronal magnetic field, which suddenly
released its free energy through eruptions. In order to understand how
the eruptions are triggered, many theoretical models have been
proposed
\citep{sakurai1976,van1989,antiochos1999,chen2000,moore2001,torok2005,kliem2006,aulanier2010},
which can roughly be divided into two categories: one is based on
magnetic reconnection, and the other is via ideal MHD process
\citep{sakurai1976,torok2005,kliem2006,aulanier2010}. The reconnection
mechanisms, including mainly the tether-cutting
model~\citep{moore2001} and break-out model~\citep{antiochos1999},
consider that magnetic reconnection plays a fundamental role in
producing the eruption of magnetic structures. The tether-cutting
model suggests that reconnection within sheared magnetic arcades can
break the balance between the outward magnetic pressure force and the
inward magnetic tension of the overlying field~\citep{moore2001}. It
describes the situation that two magnetic arcades get close to each
other via photospheric shearing and converging flows. When the sheared
arcades are close enough, magnetic reconnection will take place to
build up a magnetic flux rope (MFR) or even trigger an eruption. This
model has been well studied during last decades, as numerous
observations have shown shearing and converging flows a few hours or
minutes before triggering of flares and CMEs
\citep{wang2005,liu2012,shimizu2014}. And accompanied magnetic
cancellations and increasing radiations indicate that magnetic
reconnection takes place and finally triggers the eruptions
\citep{leka1996,sterling2010}. Note that the reconnection is within
the core structure.
On the other hand, 
the break-out model proposes that reconnection above the sheared
arcades can do the same work by weakening the overlying
field~\citep{antiochos1999}. Such process studied by simulations is also consistent with observations \citep{lynch2004}. Based on coronal magnetic field extrapolations, it is demonstrated that this model can explain observations of CMEs \citep{zhou2017}. Unlike the tether-cutting model, there are fewer precursors that can be observed for this model \citep{sterling2001a}. Besides, magnetic flux emergence and its associated magnetic reconnection are also proposed as being the trigger of eruption. For example, \citet{chen2000} proposed a model in which the emergence flux inside or outside the core structure (a MFR) can trigger its eruption, i.e., an inside emerging flux can cause converging flow underneath the MFR and then trigger reconnection, while an outside emerging flux can weaken the overlying field via reconnection. Note that all these mentioned models are based magnetic reconnection. The ideal MHD models believe that the
eruptions can be triggered and accelerated by the ideal
magnetohydrodynamics (MHD) instabilities, such as kink instability
and torus instability associated with twisted magnetic flux
rope (MFR) \citep{sakurai1976,torok2005,kliem2006,aulanier2010}. In
such models, the magnetic reconnection is triggered as a result of the
eruptions. The simulation of \citet{aulanier2010} suggests that the ideal MHD instability plays a more important role in forming CMEs. And in the investigation of the strongest flare in 24 solar cycle by \citet{yang2017}, they think the associated CME was firstly caused by the kink instability of the filament embedded in.

Although there are numerous studies on the early evolution of CMEs, only a few of them focus on quiescent filament eruption. Basing on previous studies we can know that they are not so different from the AR eruptions. They can have similar evolve processes, reach a high enough speed or associated with a double-ribbon flare. However, some studies found that the accompanied flares have some delay with the eruption onset \citep{sterling2001b}. It seems the reconnection is the result of the eruption, just like the description of those eruptions triggered by idea MHD instabilities. And \citet{zhou2006} and \citet{su2015} are confirmed these instabilities, such as kink instability and torus instability, can trigger eruptions of quiescent filaments.

Filaments, or prominences when seen above the solar limb, are
fascinating objects embedded in the solar corona. They are rougly 100
times denser than the surrounding corona but 100 times colder. In
order to sustain such heavy materials, special magnetic structures are
needed to provide upward magnetic tension force, for example, the
magnetic dips within MFRs~\citep{ks57,kr74}. Since filament eruptions
can be well observed by multi-wavelength, such as H$\alpha$, 304\AA,
193\AA and 211\AA, once the MFR holding a filament becomes unstable
and erupts, the erupting filament as observed provides a tracer of the
dynamic evolution of the underlying magnetic structure. Furthermore,
many statistical studies have shown that the occurence of filament
eruptions are strongly related with that of CMEs
\citep{gopalswamy2003,jing2004,chen2011}, and the erupting filament
materials are recognized in most cases as the bright core of typical
CMEs. Thus observation of the filament eruptions provides the key in
studying the early kinetic evolution of CMEs.

Regarding to the triggering mechanisms by either reconnection or ideal
MHD, the filament eruptions should behave rather differently in
observations. For reconnection-triggered ones, their eruptions should
be preceded by a strong enhancement of X-ray (as well as EUV)
intensity close to or on the filament spine. For ideal MHD process,
the intensity enhancement or X-ray flux increasing will occur a bit
after its eruption. Thus for typical events, the triggers can be
easily distinguished. In their evolutions, the idea MHD process can
always provide a clear trajectory of erupting filament
\citep{cheng2013}, which can be used to study its acceleration. While
for reconnection process, more precursors can be observed in
photospheric magnetic field, such as emerging parasitic poles or
conversion flows \citep{feynman1995,jing2004}. These precursors are
helpful in predicting its trigger.

The Heliophysics Event Knowledgebase \citep[HEK,][]{hurlburt2012},
which is a catalog of the records of solar activities, including the
filament eruption records from 2011 to now. Using the records, some
statistical studies discussed the relationship between eruptive
filaments and formed CMEs. For example, \citet{mccauley2015} surveyed
some properties of more than 900 filament eruptions recorded by
HEK. Among them, more than 100 limb events are analyzed for studying
their onset radial height and decay index. It is found that the onset
heights show a significant difference in different filament types. On
the other hand, the differences for decay indexes are not so
significant. The decay indexes have an average value of 1.1 for all
the filament, with the quiescent type slightly higher and the active
region type slightly lower. For the eruptive filaments, some of them
can form the CMEs and some of them failed. \citet{jing2018} studied 38
eruption events that are associated with flares stronger than M5
class. They found that all the core magnetic structures of confined
events are underneath the height with decay index equal $0.8$. But the
eruptive ones do not show a well-defined threshold, e.g., they can
occur with decay index even smaller than $0.8$. Although, it is
concluded that reconnection solely seems not able to strong enough for
ejecting a CME in the study of \citet{aulanier2010}, the statistical
result suggests that reconnection can break out the constrain of
strong overlying magnetic field, i.e., the eruptive events with decay
index smaller than 0.8 are triggered by reconnection. Thus a survey of
the triggering mechanisms of filament eruptions is required to provide
new insights for this topic.

In this paper, we carried out a statistical study to identify the
triggering mechanisms for filament eruptions with focus on those
resulting in high-speed CMEs. We will statistically analyse the
relationships among CME velocities, filament types, trigger mechanism
and some magnetic properties. The first part (Section~\ref{dam}) will
introduce the data, standards in dividing the events and methods in
calculating the magnetic properties. Second part (Section~\ref{redis})
will gather the results of the statistical properties. The last part
(Section~\ref{sum}) will summarize our findings.

\begin{table*}
\centering
\caption{The list of all events of filament eruptions.}
\begin{tabular}{ccccccccccc}
\toprule
Date& Type& AR$^a$ No.& FC$^a$& FOT$^a$& FPT$^a$& CME V (km/s)& Mechanism& OT$^a$& OH$^a$ (Mm)& DI$^a$\\
\midrule
2011/06/07& AR& AR11226& M2.5& 06:16& 06:34& 1255&  R$^a$& 06:16:46&  18.1& 0.77\\
2011/07/06& QS$^a$& /& /& /& /&  835& NR$^a$& 09:56:13& 182.0& 1.25\\
2011/08/04& AR& AR11261& M9.3& 03:41& 03:57& 1315&  R& 03:46:58&  39.3& 1.86\\
2011/08/11& AR& AR11263& C6.2& 09:34& 10:23& 1160&  R& 09:55:33&  22.6& 1.34\\
2011/09/13& IF$^a$& AR11288& C2.7& 12:03& 13:02& 1024& NR& 12:01:18&  49.8& 1.26\\
2011/10/01& AR& AR11309& B8.5& 20:30& 20:32& 1238& NR& 20:26:42&  72.5& 2.47\\  
2011/11/09& IF& AR11342& M1.1& 13:04& 13:35&  907&  R& 13:06:52&  82.9& 1.30\\
2012/01/01& QS& /&          /&     /&     /&  801& NR& 00:31:50&  70.8& 0.74\\
2012/01/19& AR& AR11402& M3.2& 13:44& 16:03& 1120&  R& 14:30:33&     /&    /\\
2012/01/23& IF& AR11402& M8.7& 03:38& 03:58& 2175&  R$^b$& 03:37:56&  71.7& 1.46\\
2012/01/27& AR& AR11402& X1.7& 17:37& 18:36& 2508& NR$^c$& 17:44:13&  66.7& 1.56\\
2012/02/19& QS& /&          /&     /&     /&  846& NR& 16:33:59&  42.9& 1.68\\
2012/03/16& QS& /&          /&     /&     /&  862& NR& 19:02:40& 344.1& 1.69\\
2012/03/27& AR& AR11444& C5.3& 02:50& 03:08& 1148&  R& 02:50:44&  41.7& 1.91\\
2012/04/05& AR& AR11450& C1.5& 20:49& 21:10&  828& NR& 20:48:47&  87.2& 1.92\\
2012/04/09& QS& /&          /&     /&     /&  921& NR& 14:28:27& 159.6& 1.05\\
2012/04/16& AR& AR11461& M1.7& 17:24& 17:40& 1348&  R& 17:24:32&  17.0& 2.07\\
2012/05/11& QS& /&       C3.2& 23:02& 23:44&  805&  R& 23:17:13&     /&    /\\
2012/06/23& AR& AR11506& C2.7& 07:02& 07:46& 1263&  R& 06:58:32& 115.8& 0.78\\
2012/07/02& QS& /&          /&     /&     /&  988& NR& 06:12:43& 129.3& 1.69\\
2012/07/12& QS& /&          /&     /&     /&  843& NR& 14:50:38&  39.0& 0.84\\
2012/08/17& IF& AR11542& B6.0& 21:51& 22:28&  986&  R& 21:52:43&     /&    /\\
2012/08/31& IF& AR11562& C8.5& 19:33& 20:43& 1442& NR& 19:14:06&  80.2& 2.20\\
2012/09/27& IF& AR11577& C3.7& 23:28& 23:57&  947& NR& 23:19:31&  56.3& 0.76\\
2012/11/27& QS& /&       C5.9& 02:10& 02:15&  844& NR& 01:16:34&  75.3& 0.85\\
2012/12/04& IF& AR11628& C1.8& 00:05& 00:28&  963& NR& 23:49:43&  98.9& 0.88\\
2013/01/06& QS& /&          /&     /&     /&  860& NR& 05:37:35& 103.8& 0.88\\
2013/02/06& AR& AR11667& C8.7& 00:04& 00:20& 1867&  R& 00:09:11&     /&    /\\
2013/02/11& AR& AR11667& B5.8& 18:56& 19:05& 1161& NR& 18:53:57&  19.6& 1.90\\
2013/02/12& QS& /&          /&     /&     /& 1050& NR& 21:42:07&  38.0& 2.08\\
2013/03/12& AR& AR11690& C2.0& 10:17& 10:39& 1024& NR& 10:16:23&  94.1& 2.11\\
2013/07/18& QS& /&       C2.3& 19:57& 20:14&  939& NR& 17:52:03&  68.1& 1.58\\
2013/08/17& AR& AR11818& M3.3& 18:16& 18:24& 1202&  R& 18:21:36&  19.3& 1.31\\
2013/09/10& QS& /&          /&     /&     /&  847& NR& 11:59:04&  32.5& 0.66\\
2013/09/24& QS& /&          /&     /&     /&  919& NR& 20:03:25& 136.8& 0.69\\
2013/09/29& QS& /&       C1.2& 21:43& 23:30& 1179& NR& 21:23:55&  51.2& 0.89\\
2013/10/28& AR& AR11875& M5.1& 04:32& 04:41& 1201&  R& 04:38:10&  24.2& 0.59\\
2013/10/28& AR& AR11882& M4.4& 15:07& 15:15&  812& R$^d$& 14:54:31&     /&    /\\
2013/12/23& QS& /&          /&     /&     /& 1409& NR& 07:47:07&  91.7& /$^e$\\
2014/01/03& QS& /&          /&     /&     /& 1132& NR& 03:11:26&  46.5& 2.41\\
2014/01/30& AR& AR11967& M6.6& 15:52& 16:11& 1087& NR& 15:49:27&  44.0& 1.50\\
2014/02/25& AR& AR11990& X4.9& 00:39& 00:49& 2147&  R& 00:41:57&  17.8& 0.40\\
2014/03/04& QS& /&          /&     /&     /&  911& NR& 20:58:24&  20.0& 1.53\\
2014/04/02& AR& AR12027& M6.5& 13:18& 14:03& 1471& NR& 13:15:45&  61.3& 1.70\\
2014/04/12& AR& AR12035& C5.0& 07:04& 07:31& 1016&  R& 07:14:21&  40.2& 1.52\\
2014/04/18& QS& /& /$^f$&     /&     /&  926&  R& 08:54:19&    /&    /\\
2014/06/08& AR& AR12087& C5.2& 23:46& 23:54&  836&  R& 23:47:51&  11.5& 0.81\\
2014/06/13& QS& /& /$^g$&     /&     /&  992& NR& 18:11:04& 119.0& 0.90\\
2014/07/10& QS& /&          /&     /&     /&  928& NR& 05:57:37& 157.0& 1.75\\
2014/12/12& IF& AR12232& C1.2& 03:42& 04:51& 1133& NR& 03:22:00&     /&    /\\
\bottomrule
\end{tabular}
\label{tab0}
\end{table*}

\begin{table*}
\centering
\caption{Continued of Table~1.}
\begin{tabular}{ccccccccccc}
\toprule
Date& Type& AR$^a$ No.& FC$^a$& FOT$^a$& FPT$^a$& CME V (km/s)& Mechanism& OT$^a$& OH$^a$ (Mm)& DI$^a$\\
\midrule
2014/12/20& QS& /&          /&     /&     /&  830&  R& 00:58:01&  26.9& 0.26\\
2015/01/12& AR& AR12261& C3.7& 15:18& 15:21& 1078& NR& 15:10:09&  53.3& 1.05\\
2015/04/04& QS& /&       C3.8& 22:16& 23:48&  825&  R& 22:20:31&  69.9& 0.64\\
2015/06/03& QS& /& /$^h$&     /&     /&  818& NR& 22:20:31& 118.9& 1.34\\
2015/06/18& AR& AR12365& M1.3& 00:20& 01:27& 1714&  R& 00:20:34&  24.6& 1.73\\
2015/07/25& QS& /&          /&     /&     /&  889& NR& 13:32:30&  64.6& 1.56\\
2015/08/07& QS& /&          /&     /&     /&  876& NR& 19:27:10& 167.1& 1.02\\
2015/08/09& IF& AR12393& B8.1& 11:03& 11:43&  810& NR& 10:51:06&  54.4& 1.93\\
2015/08/23& QS& /&       C1.4& 04:07& 04:22&  832&  R& 04:08:06&     /&    /\\
2015/10/07& QS& /&          /&     /&     /&  900& NR& 07:12:15& 174.6& 0.95\\
2016/04/18& QS& /& /$^g$&     /&     /&  822& NR& 01:38:09&  29.9& 1.27\\
2016/05/11& IF& AR12540& /$^h$& /&   /&  967& NR& 10:34:41&  87.1& 1.34\\
2016/05/24& QS& /&          /&     /&     /&  914& NR& 04:16:57&  75.0& 0.96\\
2017/04/17& IF& AR12651& B7.6& 21:23& 21:42& 1014&  R& 21:31:30&  38.6& 0.99\\
2017/04/20& AR& AR12652&    /&     /&     /& 1041& NR& 16:15:27&  21.7& /$^i$\\
2017/04/24& QS& /&          /&     /&     /& 1008& NR& 01:38:26&  69.8& 0.68\\
\bottomrule
\end{tabular}
  \begin{tablenotes}
  \item[a] $^a$ AR: AR filament, IF: Intermediate filament, QS: QS filament, R: reconnection, NR: non-reconnection, FC: flare class, FOT: flare onset time, FPT: flare peak time, OT: onset time, OH: onset height, DI: decay index
  \item[b] $^b$ Before the eruption and main flare, there is a weaker flare underneath the filament. After the flare ribbon propagate through the filament north part, the filament erupted. Therefore, we believe that this filament was triggered by the reconnection.
  \item[c] $^c$  Before the flare triggered, the filament show obvious kink motion, therefore this eruption is considered as a non-reconnection triggered one.
  \item[d] $^d$  At about 14:50, before the onset time of flare and eruption, the south footpoint of filament start flaring and a slow X-ray flux increase is associated. Therefore, we believe this filament eruption is triggered by this reconnection.
  \item[e] $^e$  The surrounding magnetic field is too weak to avoid the error from the noise of magnetic field strength.
  \item[f] $^f$  The eruption do associate a small flare, but we cannot determine whether the soft X-ray flux rising is caused by this flare or the decay phase of the previous flare in other ARs.
  \item[g] $^g$  The GOES flux is covered by other flares
  \item[h] $^h$  There is no GOES record.
  \item[i] $^i$  There is no PFSS data.
  \end{tablenotes}
\end{table*}

\section{Data and Method} \label{dam}

The filament eruption samples are obtained from HEK. Here we first
selected the significantly high-speed CME events (velocity
$> 800$~km~s$^{-1}$ which is larger than the speed of fast solar wind)
in time span from June 1 2011 to June 1 2017 and identified their
origin sites. Then, only the CMEs which are accompanied with filament
eruptions are confirmed as the samples. A total number of 87 events
are selected from over 400 CMEs. Among these 87 events, some of them
took place behind the solar disk and a few of them are hardly to be
traced for their trajectory. As such, we discarded these events since
they will cause errors in identifying their flare class and onset
time. After this filtering, we finally obtained 66 events for
statistic study. All of their eruption evolutions can be observed by
Atmospheric Imaging Assembly \citep[AIA,][]{lemen2012} onboard the
Solar Dynamics Observatory \citep[SDO,][]{pesnell2012}. Furthermore,
in order to gain the accurate position information of the pre-eruptive
filaments, 195~\AA\ images taken by Solar Terrestrial Relations
Observatory \citep[STEREO,][]{howard2008} are also used. All the
events are listed in Table~\ref{tab0}.

\begin{figure*}
\centering
\includegraphics[width=0.8\textwidth]{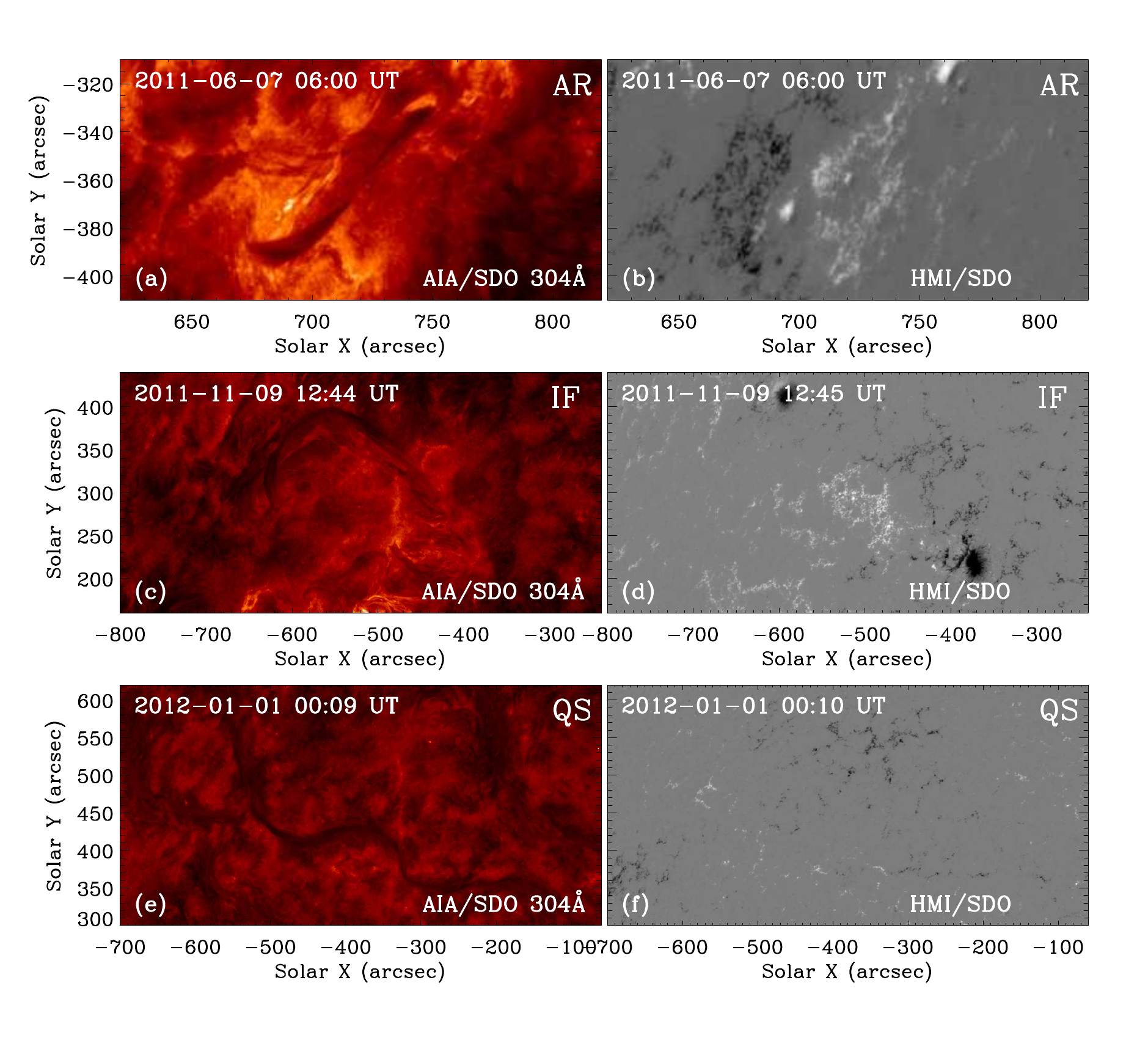}
\caption{The samples of typical filaments in different types observed
  in 304\AA\ and 193\AA\ filtergrams.}
\label{fig1}
\end{figure*}

\begin{figure}
\centering
\includegraphics[width=0.5\textwidth]{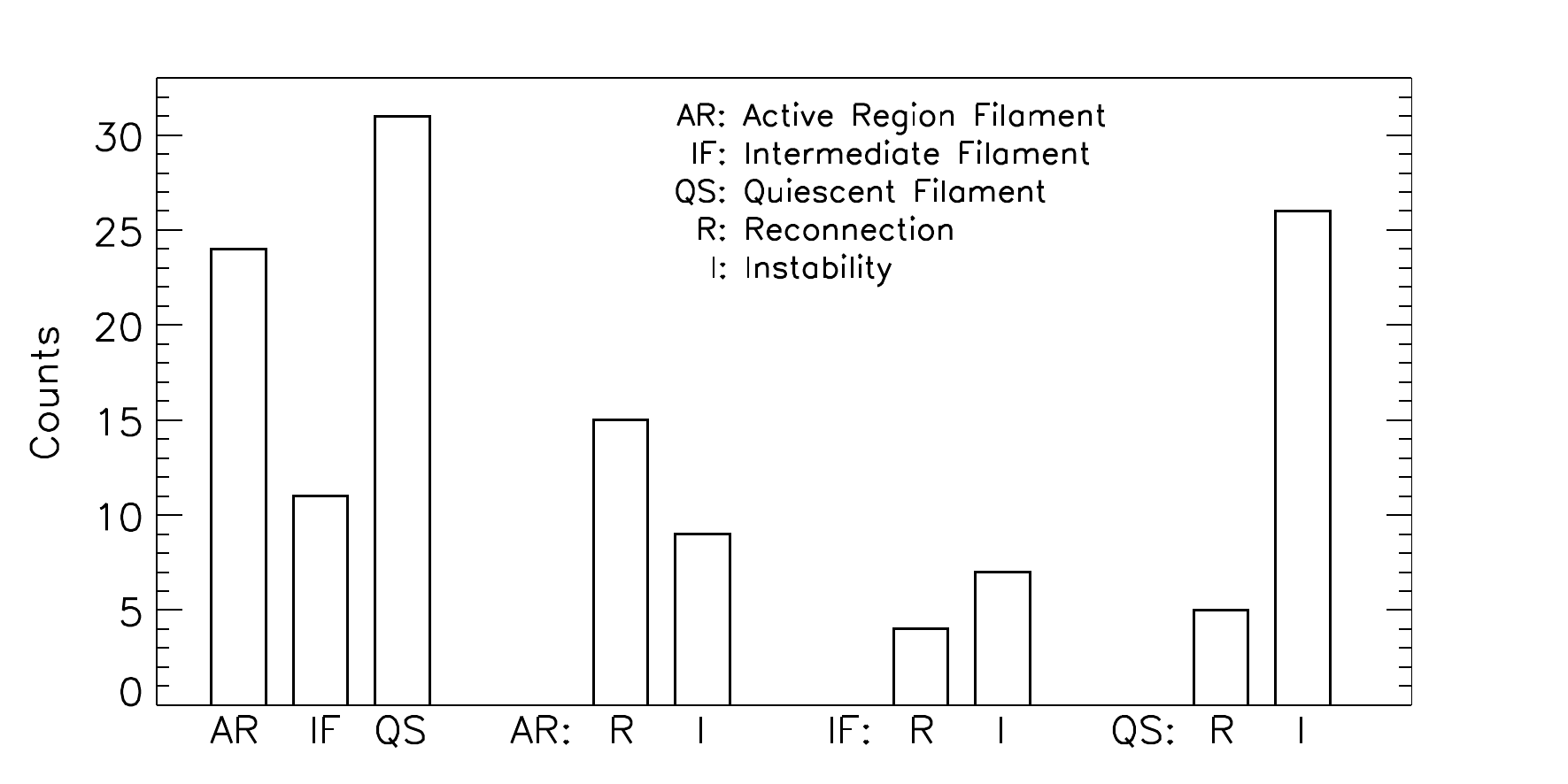}
\caption{The histogram of filament types and mechanisms of different
  types.}
\label{fig2}
\end{figure}

\subsection{Filament types}

Filaments (or prominences) distribute in the whole solar disc,
including the polar regions as well as the equator
\citep{engvold2015}. They are often grouped in four types according to
their locations. Active region (AR) filaments form in ARs. Quiet Sun
(QS) filaments are located in quiet sun regions far away from
ARs. There are also intermediate filaments (IF) that have one leg
rooted in AR and the other one rooted in quiet sun region, or lie
between neighboring ARs. Filaments found at the polar regions, e.g.,
at latitudes larger than 50$^{\circ}$, are called polar crown (PC)
filaments. Different types of filaments can have different physical
properties, e.g., magnetic configurations and strength. Thus dividing
the samples by filament types can help us clarifying the important
factors that affect their eruption. A typical filament for each type
is shown in Figure~\ref{fig1}, and the statistical numbers are given
in Figure~\ref{fig2}. Note that there is no PC filament in our
statistic samples, probably due to the fact that the weak magnetic
field strength in high latitude is unable to produce fast CMEs.

\begin{figure*}
\centering
\includegraphics[width=0.8\textwidth]{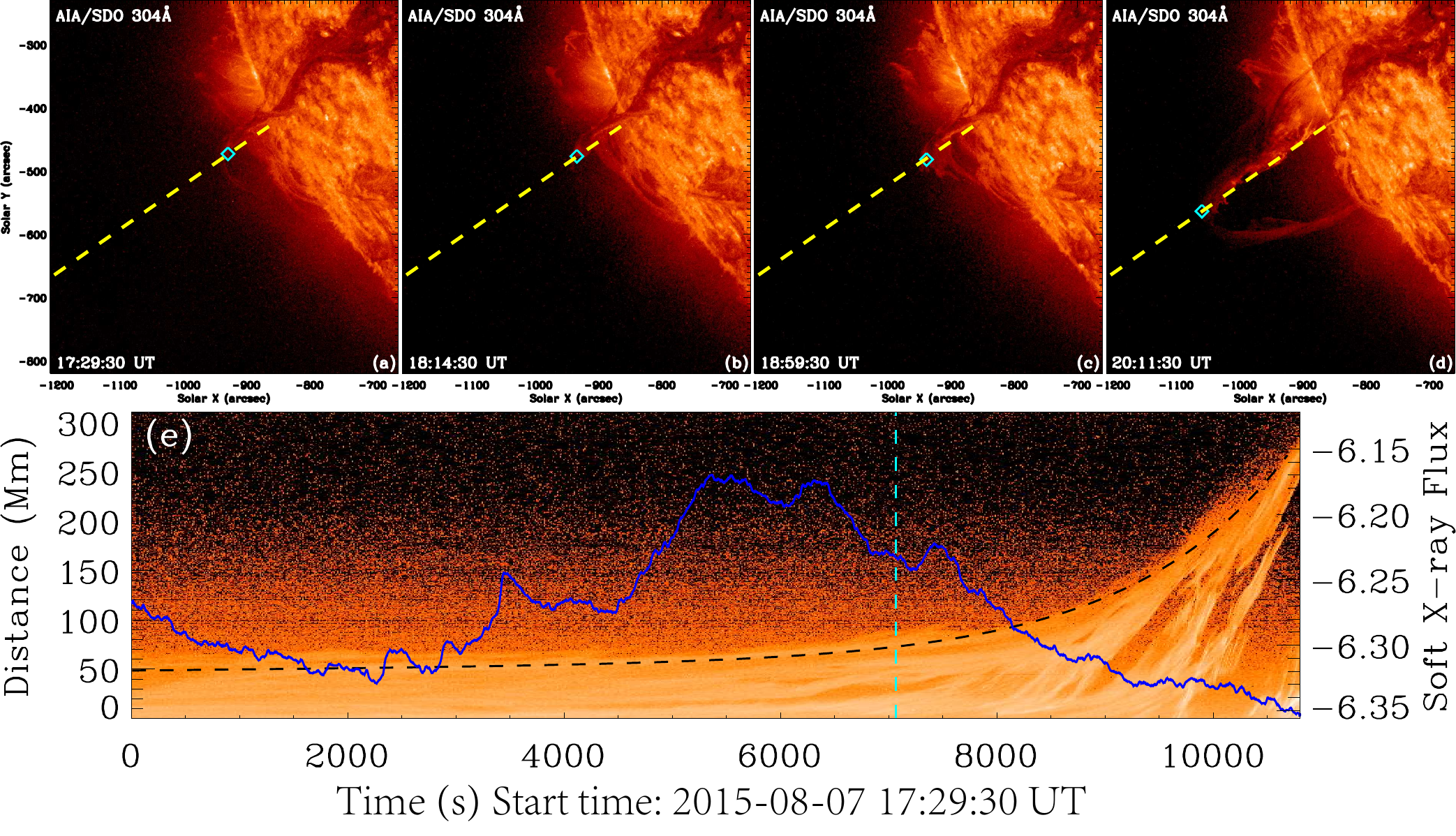}
\caption{The example for evaluating the time-distance maps. The top panels
  show the marks in cyan diamond and the slice in yellow
  dashed line. Panel e show the time-distance map of the sample. The
  blue line is GOES soft X-ray flux, the black dashed line is
  time-distance fitting curve and the cyan dashed line shows the onset
  time.}
\label{fig3}
\end{figure*}

\begin{figure*}
\centering
\includegraphics[width=0.7\textwidth]{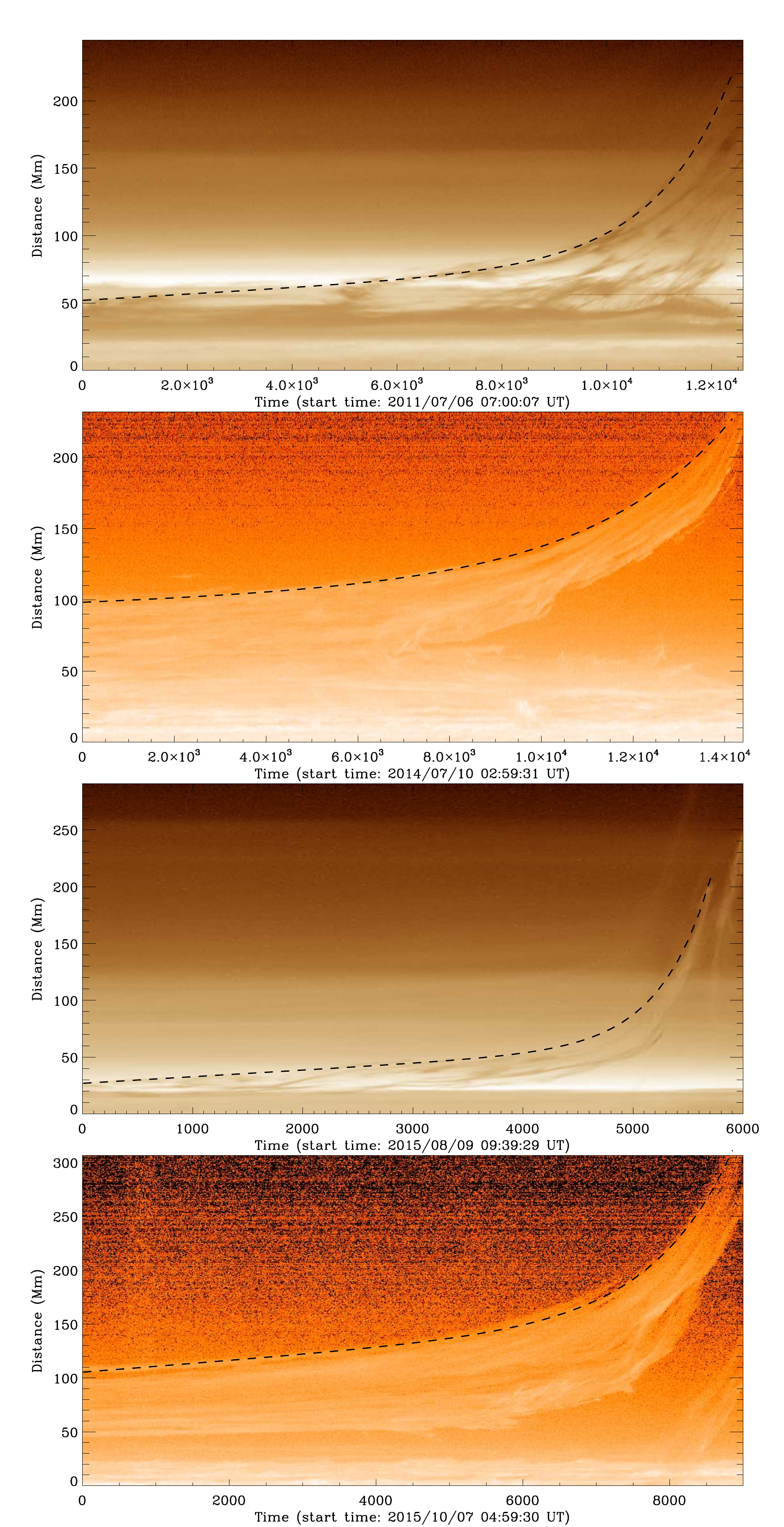}
\caption{Four well-fit examples fitted with equation~\ref{equ1}.}
\label{fig4}
\end{figure*}
\begin{figure*}
\centering
\includegraphics[width=\textwidth]{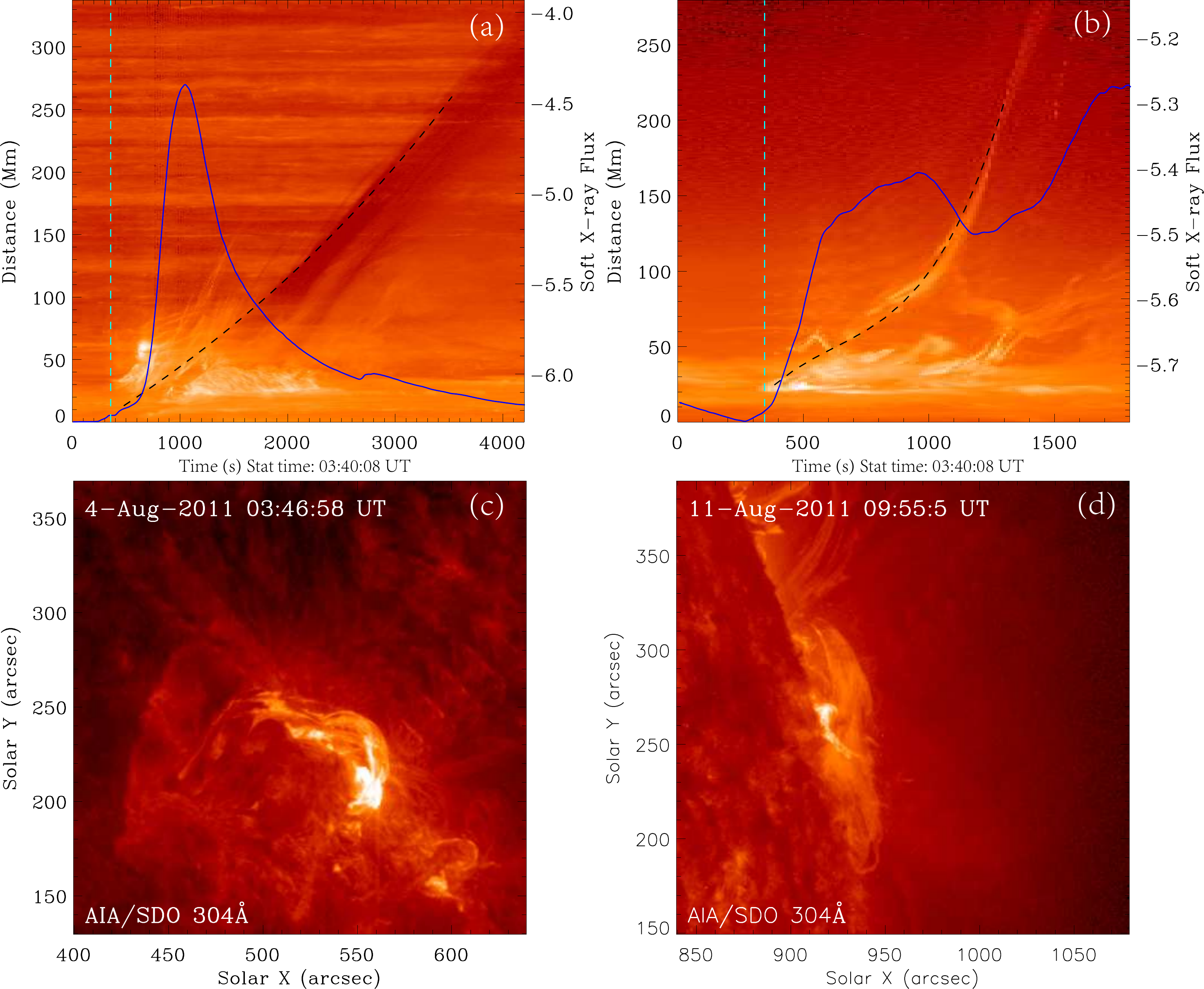}
\caption{Samples for clarifying the reconnection triggered events. The
  black dashed lines in panels (a) and (b) are time-distance
  fitting curves, the blue lines are GOES soft X-ray flux, and
    cyan dashed lines indicate the onset time. Panels (c) and (d) are
    shown for the filaments observed at the onset time respectively.}
\label{fig6}
\end{figure*}

\begin{figure*}
\centering
\includegraphics[width=0.7\textwidth]{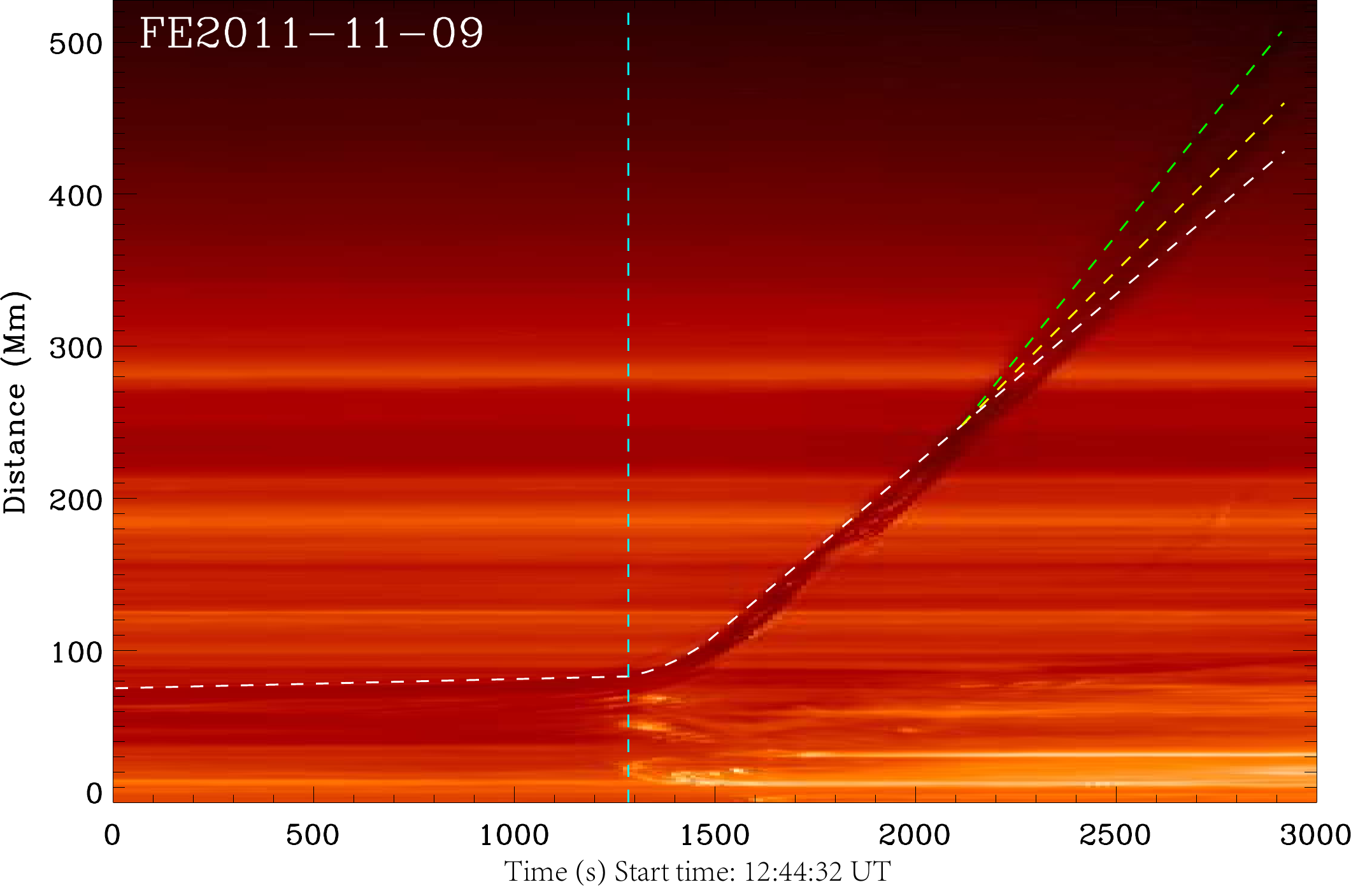}
\caption{A sample of time-distance map that cannot be fitted by
  Equation~\ref{equ1}. The white, yellow and green dashed lines are
  time-distance fitting curve. The cyan dashed line shows the
    onset time of this eruption.}
\label{fig5}
\end{figure*}

\subsection{Time-distance evolutions}

Time-distance maps provide the kinetic information of a drifting
object with time. It is widely used in tracing eruptive events, such
as jets, flares and filament eruptions. In a majority of the cases, a
straight line pointing along the eruption direction can fully cover
the evolution trajectory of an erupting filament. However, for a few
events, the trajectories are not along a straight line. In such cases,
we select 4 or 5 frames during the eruption and mark the front of
eruption by hand. Then a conic is used to fit all the mark points. In
order to cover the whole trajectory of eruptions, we extended this
conic from the footpoint to the front until it moves out of the
view. Figure~\ref{fig3} shows one of the events analyzed for time-distance evolution.
The upper panels show the selected
frames, where the marks are shown by the green diamonds. The yellow
dashed line shows the location for picking up the slice. The bottom
panel displays the time-distance map along the yellow dashed line.

Evolution of an erupting filament often demonstrates two distinct
phases: a slow-rise phase and a fast-rise phase. The slow-rise phase
can be recognized as the upward motion with a constant and small
velocity \citep{sterling2005}, while the fast-rise phase can be fitted
with an exponential function \citep{goff2005,william2005}. In order to
fit these two phases with a single curve, we use the function proposed
by \citet{cheng2013}:
\begin{equation}
h\left(t\right)=c_{0}e^{\left(t-t_{0}\right)/\tau}+c_{1}\left(t-t_{0}\right)+c_{2},
\label{equ1}
\end{equation}
where $h(t)$ is the height of filament, $t$ is the time and $c_{0}$,
$c_{1}$, $c_{2}$, $t_{0}$ and $\tau$ are free parameters. With this
function, we can calculate conveniently the eruption onset
time~\citep{mccauley2015}, which is defined by
\begin{equation}
t_{\rm onset}=\tau ln\left(c_{1}\tau/c_{0}\right)+t_{0}.
\label{equ2}
\end{equation}
It means the instantaneous velocity at the onset time is twice of the
initial value, i.e., the constant speed in the slow-rise phase.  In
the bottom panel of Figure \ref{fig3}, we denote the onset time using
a cyan dashed line. This function can well fit some of the eruption
evolves, more well-fit results are shown in Figure \ref{fig4}

However, it should be noted that not all filament eruptions
demonstrate such well-defined two-phase pattern. In our samples, about
33\% cannot be fitted with the two-phase function. For instance, some
typical reconnection-triggered eruptions do not have a slow-rise
phase, since they are accelerated rather abruptly. As an example shown
in Figure~\ref{fig6}, the initiation of such reconnection-triggered
events are accompanied with strong EUV intensity enhancement, and in
these cases, the trajectory of erupting filaments is similar, i.e.,
starting with an almost static or extremely slow rise phase, then
followed by a rapid acceleration immediately after the intensity
enhancement, and finally reaching a constant velocity. We have
attempted to define their onset time using the aforementioned method
but failed, because as shown in Figure~\ref{fig6} some of their
acceleration phases are covered by the strong radiations. Besides,
there are also some fast accelerated eruptions which cannot be fitted
well by an exponential curve. For example, see such an event shown in
Figure~\ref{fig5}. This event includes 4 phases during its
eruption. The first phase was a very slow rise phase (nearly
undetectable), then it was most probably triggered by a reconnection
(see the intensity enhancements underneath the filament). After a
short acceleration, it ejected with nearly constant speed. Several
minutes later, it seems to be accelerated again, but meanwhile was
split into three parts. Then all these parts show uniform
velocities. The cases with such complex trajectories can hardly be
defined with an onset time using the aforementioned method. Thus, for
all these cases, we define the eruption onset time as beginning of the
intensity enhancement, i.e., the same onset time of flare
reconnection. Since in these cases, the filaments are almost static
before eruption and the accelerations are so rapid, there is no
significant uncertainty brought in calculating its onset height.

\subsection{Onset height and decay index}

It has been suggested that ideal MHD instabilities play an important
role in triggering filament eruptions. In particular, the torus
instability has been proposed to be the most important one for
producing CMEs. For an ideal MFR configuration, the torus instability
is determined by the decay index, which describes the decreasing speed
(with height) of background magnetic field overlying the
MFR. Specifically, the MFR system becomes unstable when the apex of
the rope axis reaches a height where the decay index exceeds a
threshold value~\citep{torok2005,kliem2006}. Thus it is meaningful to
calculate the onset height of eruption prominence, and derive the
decay index at this height.

Joint observations of STEREO and SDO are used to obtain the onset
height. Figures \ref{fig8} and \ref{fig9} show two examples of
evaluating the onset height. First, we use a diamond to mark the body
of prominence and then trace the latitude to mark the same position of
filament in STEREO. This new mark can tell us the longitude of the
filament, and associated with the relative position of STEREO and SDO,
we can get the longitude and latitude of the prominence observed in
SDO. Thus the position and height of prominence can be
derived. However, this method can be used only for those prominences
observed in both SDO and STEREO, while some events do not have STEREO
observations with an appropriate angle. For prominences (above the
limb) of these events, we calculate their radial height projected to
the plane-to-sky. For radial eruptive filaments (onboard the disc), we
can calculate the distance between filament spine and footspoints and
project it back to 3D space. Even so, there are still 8 events (3 AR
filaments, 2 IF filaments and 3 QS filaments) for which the onset
height cannot be determined.

After evaluating the onset height of filaments, we can calculate the
decay index $n$, which is defined as
$n=-\partial{ln B}/\partial{ln h}$, at their onset height $h$. The
magnetic field $B$ is obtained by PFSS extrapolation from the
magnetogram taken by Global Oscillation Network Group
\citep[GONG,][]{harvey2011}, and only the transverse component is used
in the calculation.

\begin{figure*}
\centering
\includegraphics[width=0.8\textwidth]{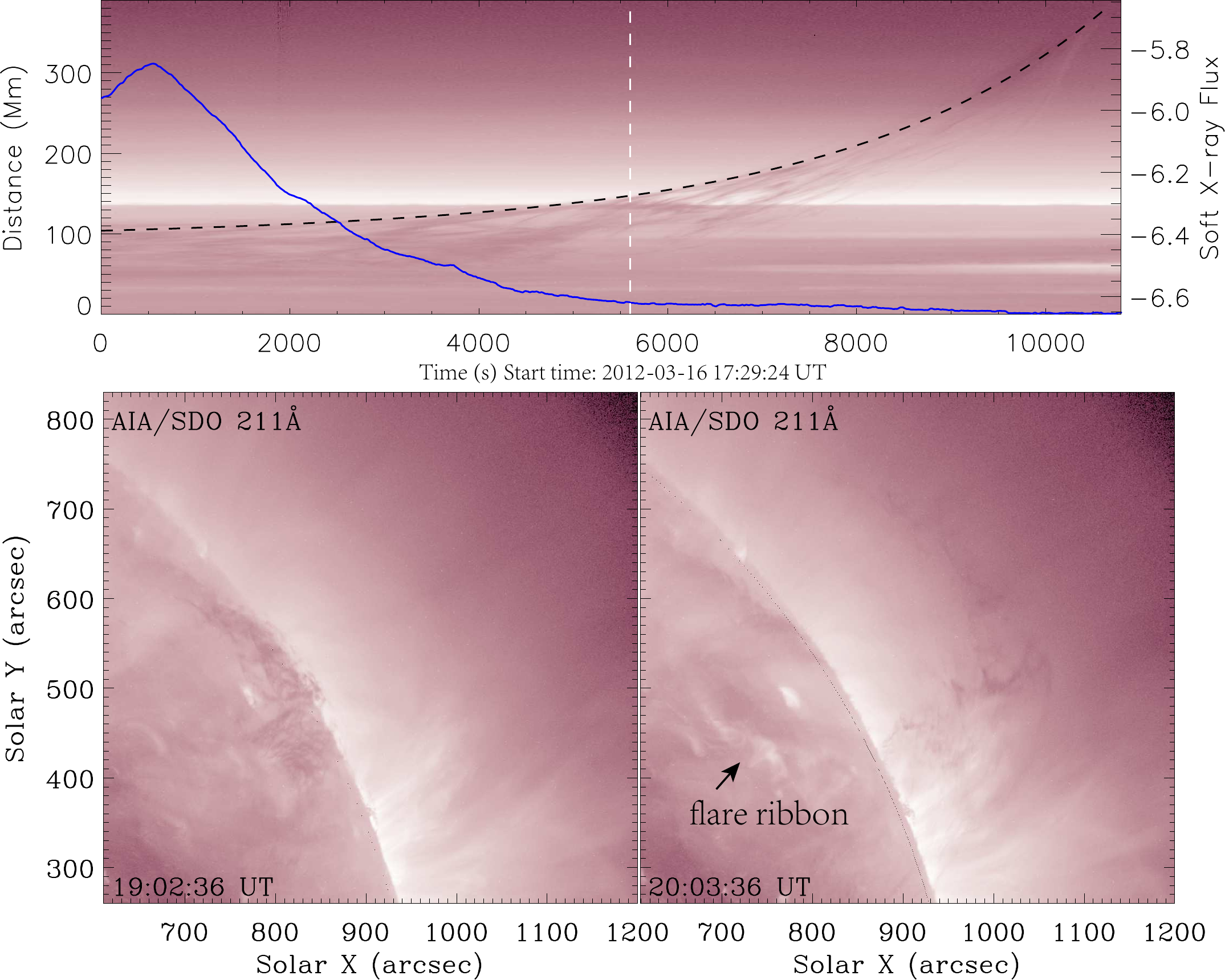}
\caption{An example of non-reconnection
  triggered events. The top panel shows the time-distance map. The blue line is GOES soft X-ray flux, black
  dashed line is time-distance fitting curve and white dashed line
  shows the onset time. The bottom left panel shows the filament position at onset
  time and the bottom right panel shows the ribbons when it can clear be seen.}
\label{fig7}
\end{figure*}

\begin{figure*}
\centering
\includegraphics[width=0.8\textwidth]{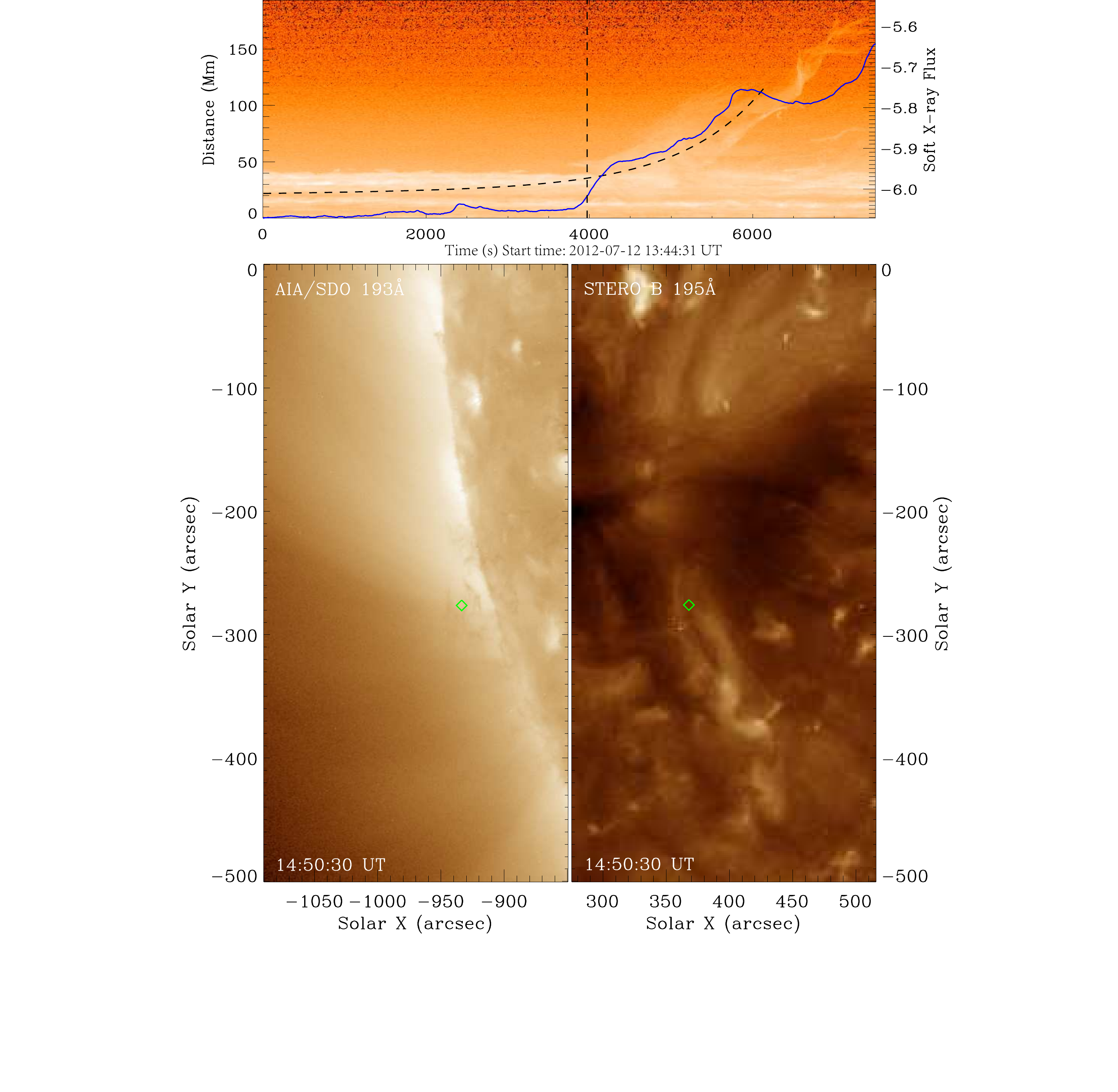}
\caption{Example of non-reconnection triggered events. The top panel shows the time-distance map of non-reconnection
  triggered sample. The blue line is GOES soft X-ray flux, black
  dashed curve is time-distance fitting curve and the vertical black
  dashed line denotes the onset time. The bottom left panel shows the filament
  position at the onset time in AIA and panel c shows it in STEREO. The
  green diamond shows the same position of filament observed in the two
  facilities.}
\label{fig8}
\end{figure*}

\begin{figure*}
\centering
\includegraphics[width=\textwidth]{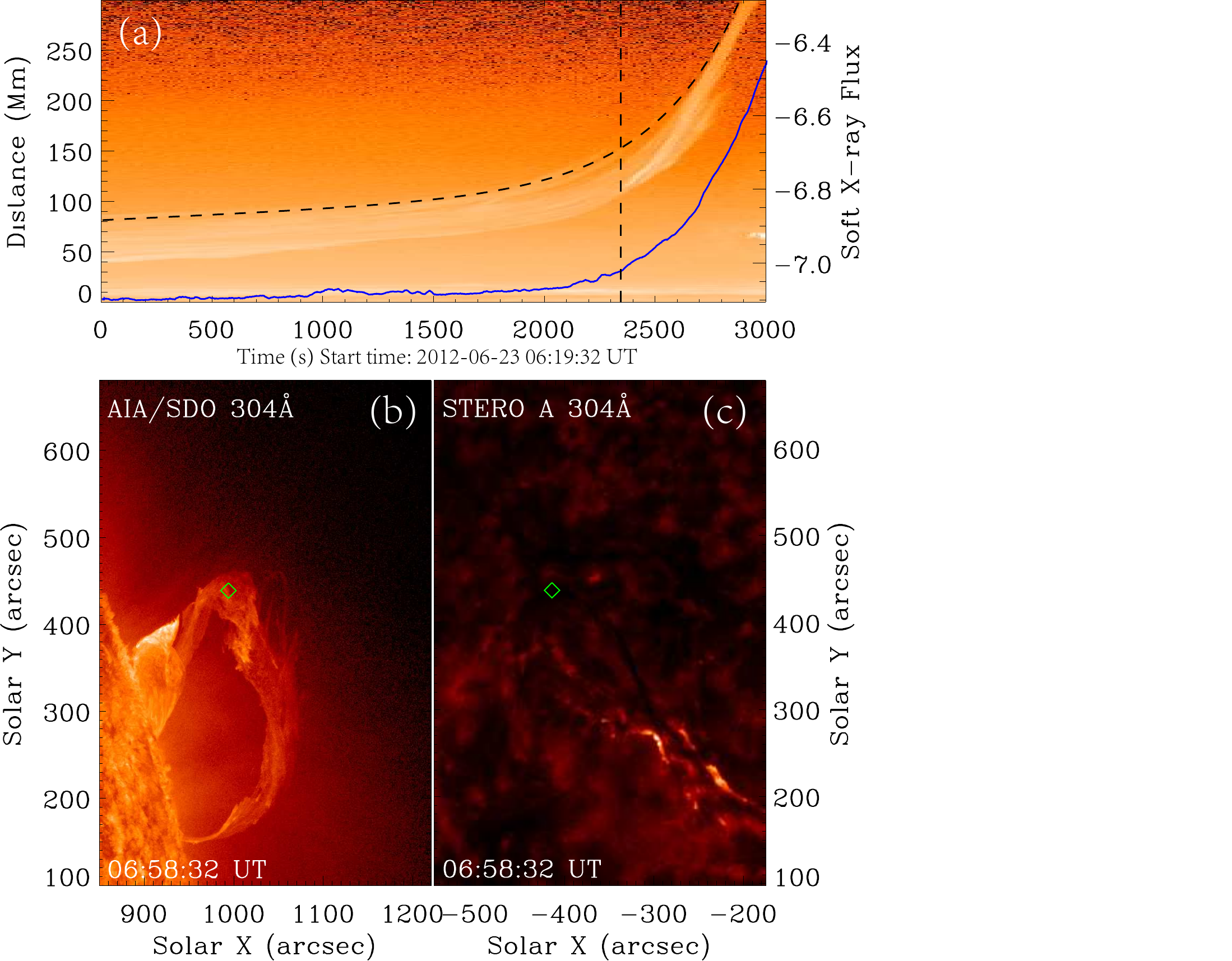}
\caption{Same as Figure~\ref{fig8} but for an reconnection-triggered event.}
\label{fig9}
\end{figure*}

\subsection{Trigger mechanism}
The trigger mechanisms of filament eruption, which have been
introduced in Section~\ref{intro}, are classified into two types:
reconnection and ideal MHD processes. Combining the soft X-ray flux,
the time-distance map and the observations at onset time, we can
distinguish which mechanism works in a filament eruption: for the
reconnection-triggered events, the increase of soft X-ray flux, which
corresponds to beginning of the associated flare, should be observed
just ahead of the eruption onset time, and newly-formed flare ribbons
or enhancements of emission intensity should be observed at the
footpoints or spine of the erupting filament. In our events, some of
them can unambiguously identified as being triggered by reconnection
because the flare emission enhancements occur clearly before the onset
of the fast rise eruption, for instance, see the sample shown in
Figure~\ref{fig5}. Two more samples of reconnection-triggered events
are shown in Figure \ref{fig6}. Along with their time-distance maps,
the observations at onset time are also exhibited. It can be seen that
at onset time, both of their X-ray flux increased and the intensity
enhancements can be seen in footpoint (panel c) and spine (panel d) of
filament. On the contrary, a typical sample of eruptions triggered by
non-reconnection processes are shown in Figure \ref{fig7}. It shows
that there is no significant X-ray flux increasing before and at onset
time and no significant radiative enhancements along the
filament. Indeed, the associated flare ribbons appear almost one hour
after the onset time of the filament eruption (see the last panel).

There are also some events which are difficult to determine the
trigger mechanism using the above method, and we give cautions for
such cases. For example, in Figure \ref{fig8}, we present a QS
prominence eruption event which appears to be triggered by
reconnection but is actually not. The time-distance map shows that the
GOES soft X-ray flux increased slightly before the eruption onset time
(which is denoted by the vertical dashed line in panel a). However, by
checking the 195 \AA\ images taken by STEREO B, we find no flare
ribbons or even no intensity enhancements associated with this
eruption. We thus further examined the causes of the X-ray flux
increasing and found that it was actually caused by an X1.4 flare
occurs at an AR, NOAA 11520. Thus this event should be considered as
triggered by ideal MHD process. For comparison, we also show a sample
which have a similar situation, i.e., there is a small GOES emission increasing before the onset time of
eruption (see Figure \ref{fig9}). At this time, there are no flares
occurred in other active regions. Furthermore, from the observation of
STEREO A, we find that underneath the erupting filament, there can
obviously be seen the newly formed flare ribbons, suggestion that this flare is accompanied with the
filament eruption. Thus this eruption can be considered as being caused by reconnection.

However, some studies suggest that although the reconnection is
observed prior to the onset time, it is still hard to say which
mechanism is the trigger as the signatures of reconnection often
coincides very closely to the onset of fast rising phase. Furthermore,
even the reconnection begins well before the eruption, there is doubt
that such reconnection is not a trigger, but an approach gradually
building up the MFR to the state of loss of equilibrium or MHD
instability~\citep{vrsnak1990,vrsnak2008,aulanier2010}. Thus we
checked our reconnection-triggered events with more evidences by
studying the decay index as well as the magnetic topology of their
soure regions. Firstly, the events for which the onset heights are far
below the torus instability threshold are not likely triggered by
torus instability. For the other events whose decay indexes are close
to or even above the torus instability threshold (as listed in
Table~\ref{tab0}), the uncertainty on the trigger mechanism might be
large since torus instability is also an option. However, as shown by
their time-distance maps (e.g., Figure~\ref{fig5}), we find the
filaments are almost static without any rising motion before the
reconnection occurs, and then they are strongly accelerated
immediately after the reconnection onset. This means the reconnection
is the key factor in breaking the equilibriums. Thus in such cases,
{it is more reasonable to consider that the eruptions are triggered by
  reconnection rather than torus instability.}
Furthermore, we checked the magnetic environments of the source region
of these eruptions by analysing the photospheric magetic flux
distribution. It is believed that reconnection-triggered event is more
likely to occur in complex magnetic topology than in the simple bipole
field. For example, \citet{moore2006} present three scenarios for
reconnection-triggered eruptions under quadrupolar configuration.
Consistently, we find that in our studied events, more than two thirds
of the reconnection-triggered ones occurred in complex magnetic
environments, e.g., a multipolar magnetic configuration. While for the
non-reconnection triggered ones, more than two thirds occurred in
simpler environment such as bipole field or the edge of ARs.
{Finally, it should be noted that there are events in which
  both reconnection and torus instability contribute to produce the
  eruption, as the reconnection first plays a role of trigger and
  shortly afterwards the torus instability sets in and further
  accelerates the filament.}

\begin{table*} 
\centering
\caption{The CME velocities of different sets. (Unit: km s$^{-1}$)}
\begin{tabular}{ccccc}
\toprule
Type$^{*}$&          Average velocity&   SE$^{*}$&    Max. velocity&    Min. velocity\\
\midrule
           AR&                   1285.00&         83.10&             2508&              812\\
          AR/R&                   1293.60&        93.71&             2147&              812\\
         AR/NR&1270.67(1116.00)$^{**}$& 175.21(65.77)&       2508(1471)&              828\\
           IF&                   1124.36&        115.78&             2175&              810\\
          IF/R&           1270.50(969.00)& 302.35(32.04)&       2175(1014)&              907\\
         IF/NR&                   1040.86&        76.11&             1442&              810\\
           QS&                    921.00&         23.15&             1409&              801\\
          QS/R&                    843.60&         21.15&              926&              805\\
         QS/NR&                    932.04&        26.83&             1409&              801\\
            R&                   1196.00&         82.27&             2175&              805\\
           NR&                   1022.74&         44.73&             2508&              801\\
            M&                   1300.58&        105.19&             2175&              812\\
            C&                   1057.65&         57.13&             1867&              805\\
           UC&                    955.18&         26.59&             1409&              801\\
\bottomrule
\end{tabular}
  \begin{tablenotes}
  \item[*] * AR: AR filament, IF: Intermediate filament, QS: QS filament, R: reconnection, NR: non-reconnection, M: M-class flare, C: C-class flare, UC: under C-class flare and SE: standard error.
  \item[**] ** bracket: the statistical results ignored the abnormal event.
  \end{tablenotes}
\label{tab1}
\end{table*}

\section{Results and Discussion} \label{redis}

Here we summarize the results found basing on the data and analysis
methods described in Section~\ref{dam}:

First, as shown in Figure~\ref{fig2}, our statistic events includes 24
AR filaments, 11 IFs and 31 QS filaments but no PC filaments. In the
statistical analysis of \citet{mccauley2015}, they mentioned that
there are 19.6\% AR filaments, 14.1\% IFs, 51.3\% QS filaments and
15\% PC filaments. The difference in type ratios of our events with
theirs suggests that high-speed CMEs result from AR filaments and IFs
more frequently than QS and PC filaments. This is rather natural since
the fast CMEs need more free magnetic energy (and stronger magnetic
field) for large acceleration. It is worthy noting that there are
still near 50\% fast CMEs formed by eruptive QS filaments, which
is more than that results from AR filaments. This indicates that the
conversion ratio of free magnetic energy to CME's kinetic energy also
plays an important role in determining CME acceleration. Such a
conversion rate is probably related to magnetic environment of the
eruptive filament. However, to determine this factor, one needs more
comparative studies between eruptive AR filaments and QS filaments.

Second, as shown again in Figure \ref{fig2}, the eruptions triggered
by reconnection are 62.5\% for AR filaments, 36.4\% for IFs and 16.1\%
for QS filaments. It shows that AR filament eruptions are preferred to
be triggered by reconnection. The IF eruptions have a similar ratio
between the reconnection and ideal-MHD triggers. On the other hand,
the majority of the QS filament eruptions are triggered by ideal MHD
mechanism, with only 5 events triggered by reconnection. It is
commonly believed that due to the weak magnetic field, reconnection in
QS area can hardly trigger a typical QS filament eruption with a large
amount of mass. Thus we carefully check the movie of these 5 events to
see why they are triggered by reconnection. We find that four of them
(FE2012-05-11, FE2014-04-18, FE2015-04-04 and FE2015-08-23) are
located at or close to the small poles which are not recognized as an
active region, and the filaments are small as typical AR ones. It
means they have a similar magnetic environment with relatively small
AR filaments, thus the reconnection process are
reasonable. Furthermore, the remaining one of the 5 events is
triggered by an X-class flare in an AR which shows a very strange
flare ribbons (FE2014-12-20). Its ribbons extend out of the AR and
trigger the reconnection far away from the AR, then lead to the
eruption of the QS filament. This is probably due to the complex
magnetic environment there.

Third, the velocities of CMEs behave very differently between
different filament types. As shown in Table~\ref{tab1}, the average
velocity of AR-filament formed CMEs is 1285.00~km~s$^{-1}$, which is
about 1.14 times of IF eruptions and 1.39 times of QS filament
eruptions. This result for the fast CMEs differs significantly from
the one summarized for all eruptions by~\citet{mccauley2015}. Unlike
their results, the fast CMEs are prefer to be formed by AR filaments
but not IFs. It means that magnetic strength plays an important role
in acceleration of CME for the formation of fast CMEs, which is in
agreement with the studies of some authors
\citep{qiu2005,chen2006}. On the other hand, the trigger mechanisms,
which are supposed to determine the initial acceleration of CMEs, i.e., the reconnection-triggered events always experience a impulsive and short acceleration phase and then evolve with a stable velocity, while the non-reconnection ones often show an exponential rising profile as described by Equation~\ref{equ1}, play
a less important role in determining the final velocity of CMEs. {As
can be seen, for whole data set, the average velocity of reconnection-triggered events is
1196.00~km s$^{-1}$ and for non-reconnection events is 1022.74~km
s$^{-1}$, respectively. Interestingly, the average CMEs velocity of
reconnection-triggered QS filament eruptions (843.60~km s$^{-1}$), is obviously smaller than
those triggered by non-reconnection processes (932.04~km s$^{-1}$).} This indicates that in
weaker magnetic environment, the efficiency of acceleration via ideal
MHD instability is higher than that via reconnection. For QS filaments
and IFs, it seems the average velocities also have a big difference
between reconnection triggered and non-reconnection triggered. But we
find that there is one sample that diverges obviously from other
samples in both two sets (FE2012-01-23 and FE2012-01-27), thus we also
show the average velocities excluding these events in Table~\ref{tab1}
with brackets. After excluding these two events, the difference is
significantly reduced.

In previous works, the class of associated flares are also considered
to be an important factor in accelerating the CME velocity
\citep{hundhausen1997,yashiro2002,vrsnak2005}. Here we also analyse
the relationship between CME velocities and associated flares. There
are only 2 X-class flares and they are associated with the CMEs with
highest velocities of above 2000 km~s$^{-1}$. There are 12 M-class
flares events, and 10 are associated with AR filaments eruptions and 2
with IFs. There are 20 events associated with C-class flares, 9 of
them are AR filament eruption, 5 of them are IFs and 6 of them are QS
events. 27 events of the rest are associated with B-class flares or
even weaker flare signal, 2 events have no record of GOES
flux and 3 events covered by other flares. The average velocity of
CMEs associated is 1300.58 km s$^{-1}$ for M-class flares, 1046.42
km s$^{-1}$ for C-class, and 956.08 km s$^{-1}$ for B-class and below.
Previously, \citet{vrsnak2005} noticed that the CMEs associated with larger flares are generally faster and broader than those with small flares. Similar result is also found by \citet{bein2012}. Here we provide a more detailed evidence.

\begin{table*}
\centering
\caption{The onset height and decay index.}
\begin{tabular}{ccccccccc}
\toprule
Type$^{*}$& Onset height (Mm)& SE$^{*}$ (Mm)& Max. h (Mm)& Min. h (Mm)& Decay index&    SE& Max DI$^{*}$& Min DI\\
\midrule
           AR&             43.44&             6.43&       115.8&        11.5&        1.45&  0.13&            2.47&    0.4\\
          AR/R&             32.66&             8.09&       115.8&        11.5&        1.26&  0.17&            2.07&    0.4\\
         AR/NR&             57.82&             6.43&        94.1&        19.6&        1.74&  0.13&            2.47&   1.05\\
           IF&             68.88&            6.68&       98.9&        38.6&        1.35&  0.16&            2.20&   0.76\\
          IF/R&             64.40&            13.30&       82.9&        38.6&        1.25&  0.14&            1.46&   0.99\\
         IF/NR&             71.12&            8.29&       98.9&        49.8&        1.40&  0.23&            2.20&   0.76\\
           QS&             100.29&            13.78&       344.1&        20.0&        1.24&  0.10&             2.41&   0.66\\
\bottomrule
\end{tabular}
  \begin{tablenotes}
  \item[*] * AR: AR filament, IF: Intermediate filament, QS: QS filament, SE: standard errors and DI: decay index.
  \end{tablenotes}
  \label{tab2}
\end{table*}


Last, the onset height and decay index are shown in
Table~\ref{tab2}. In order to clarify the effects of trigger
mechanisms, we also divide them in two sets, i.e., reconnection and
non-reconnection. Since there are only 2 events of
reconnection-triggered QS filaments for which the onset height can be
evaluated, thus they are ignored. It is found that the onset heights
of AR filaments are close to the results of \citet{mccauley2015}, but
for the IFs and QS filaments, they are a bit higher. Interestingly,
the decay indexes for different types behaves oppositely from
\citet{mccauley2015}'s results, i.e., here the values for the AR
filaments and IFs are higher while for the QS filaments they are close
to \citet{mccauley2015}'s results. The higher decay index and the same
onset height for AR filaments seems to imply that torus instability
play a more important role in accelerating faster CMEs than in
triggering the eruption. Furthermore, the average onset height (and
also the decay index) of reconnection triggered AR filaments is a bit
lower than that of non-reconnection triggered ones. However, the most
significant difference from \citet{mccauley2015}'s results is that the
minimum decay index of reconnection-triggered ones are obviously
lower. Indeed, we find the minimum decay index of non-reconnection
samples are higher than 1, but for reconnection ones, the minimum is
$0.4$. Interestingly, the eruption with such low decay index is
associated with a rather strong flare, i.e., an X4.9 class
flare. Also, there are 36\% samples have a decay index smaller than
1. We think it may explain the results summarized by \citet{jing2018},
i.e., strong enough reconnection can help magnetic structures break
out the constrain of back ground field. But these differences are not
found in IFs. Anyway, one should bear in mind that both of the onset
heights and decay indexes have big uncertainties from direct
observations. For prominences, we can get the onset heights reliably
but to calculate the decay index, we can only use the magnetogram
observed days before or after the eruption when the source region is
observed on the solar disk. Similarly, for filaments, although the
magnetogram are more accurate, the onset heights are less
certain. Thus from this result we cannot make any strong conclusion,
and a more meticulous work with more samples is needed.

\section{Summary} \label{sum}

In this paper, by joint observations of SDO, STEREO GONG and GOES, we
surveyed the eruption trigger mechanism for filaments that produced
fast CMEs with velocity above 800~km~s$^{-1}$ observed from 2011 to 2017.
The results of our statistical analysis can be summarized as follows:

(1) AR filament and IF eruptions show a higher probability for producing fast CMEs than QS filaments, while there is no PC filament eruptions producing fast CMEs in our statistic. This imply that stronger magnetic field can generate a fast CME more easily.

(2) The trigger mechanism show a significant difference between AR and
QS filaments, i.e., AR filament eruptions prefer to be triggered by
reconnection while QS filament eruptions prefer to be triggered by non-reconnection
process.

(3) The trigger mechanisms do not result in a significant difference
in the CME speeds, except for QS filament eruptions, in which non-reconnection
process can more likely generate a faster CME.

(4) Our statistic shows a trend that stronger flares are more likely associated with faster CMEs. This result is well in agreement with the previous results.

(5) The onset heights and decay indexes show differences from
previous statistical works. It shows that torus instability may {contribute} more in accelerating
faster CMEs. We also note that strong enough reconnection can
break out the constrain of background field. However, due to the uncertainty of computing
the onset heights and decay indexes, further works are needed for reconfirm this
conclusion.

In summary, we find that QS filament eruptions still have a big enough
ratio in leading to a faster CME, and they can transport free magnetic
energy into kinetic energy of CMEs with higher efficiency. Thus, it is
necessary to pay more attention in studying activities located in QS
area for space weather forecast. The trigger mechanisms play further
important roles in eruptive events. They show different behaviors in
QS and AR filaments. For predicting an AR filament eruption, more
attentions should be paid on the reconnection precursors, since the
reconnection can trigger {a} fast eruption regardless the twist
number \citep{jing2018} or even overlying magnetic field. But for QS
filaments, more attention on overlying constrained field should be
paid, since few reconnection events can lead a faster CME.

\acknowledgments 

This work is supported by the National Natural Science Foundation of
China (NSFC 41822404, 41731067, 41574170, 41531073), the Fundamental
Research Funds for the Central Universities (Grant
No.HIT.BRETIV.201901). P.Z. also acknowledges the support by China
Postdoctoral Science Foundation (2018M641812). Data from observations
are courtesy of NASA {SDO}/AIA and the HMI science teams.

\end{document}